

\input harvmac


\overfullrule=0pt


\def\T{{\scriptscriptstyle T}}


\def\CL{{\cal L}}


\def\b{\beta}

\def\e{\epsilon}
\def\g{\gamma}
\def\l{\lambda}

\def\r{\rho}
\def\u{\mu}
\def\v{\nu}


\def\aS{\alpha_s}
\def\bar#1{\overline{#1}}


\def\GeV{{\>\, \rm GeV}}
\def\gfive{\gamma^5}
\def\gtap{\raise.3ex\hbox{$>$\kern-.75em\lower1ex\hbox{$\sim$}}}
\def\Leff{\CL_{\rm eff}}
\def\ltap{\raise.3ex\hbox{$<$\kern-.75em\lower1ex\hbox{$\sim$}}}

\def\mtsq{m_t^2}
\def\muep{{\mu^\epsilon}}
\def\muepg{{\mu^{\epsilon/2} g}}
\def\Mz{M_z}
\def\nf{n_f}
\def\qbar{\overline{q}}
\def\QCD{{\scriptscriptstyle QCD}}
\def\shat{\hat{s}}

\def\space{\>\>}
\def\TeV{{\>\, \rm TeV}}
\def\that{\hat{t}}
\def\uhat{\hat{u}}


\def\sixth{{ 1\over 6}}


\newdimen\pmboffset
\pmboffset 0.022em
\def\oldpmb#1{\setbox0=\hbox{#1}%
 \copy0\kern-\wd0
 \kern\pmboffset\raise 1.732\pmboffset\copy0\kern-\wd0
 \kern\pmboffset\box0}


%

\nref\CDF{CDF Collaboration, F. Abe {\it et al.}, FERMILAB-PUB-91/231-E\semi
CDF Collaboration, F. Abe {\it et al.}, Phys. Rev. Lett. {\bf 68}
 (1992) 1104.}
\nref\Eichten{E. Eichten, K. Lane and M. Peskin, Phys. Rev. Lett. {\bf 50}
 (1983) 811.}
\nref\SimmonsI{E.H. Simmons, Phys. Lett. {\bf B226} (1989) 132.}
\nref\SimmonsII{E.H. Simmons, Phys. Lett. {\bf B246} (1990) 471.}
\nref\Politzer{H.D. Politzer, Nucl. Phys. {\bf B172} (1980) 349.}
\nref\EHLQ{E. Eichten, I. Hinchliffe, K. Lane and C. Quigg, Rev. Mod. Phys.
 {\bf 56} (1984) 579; (E) Rev. Mod. Phys. {\bf 58} (1986) 1065.}
\nref\Narison{S. Narison and R. Tarrach, Phys. Lett. {\bf B125} (1983)
 217.}
\nref\Morozov{A.Y. Morozov, Sov. J. Nucl. Phys. {\bf 40} (1984) 505.}
\nref\Braaten{E. Braaten, C.S. Li and T.C. Yuan, Phys. Rev. {\bf D42}
 (1990) 276; Phys. Rev. Lett. {\bf 64} (1990) 1709.}
\nref\Gross{D.J. Gross and F. Wilczek, Phys. Rev. Lett. {\bf 30} (1973)
  1343\semi
  H.D. Politzer, Phys. Rev. Lett. {\bf 30} (1973) 1346.}
\nref\Bethke{S. Bethke, Talk presented at XXVI Int. Conf. High Energy
 Physics, Dallas, Aug. 6-12, 1992.}
\nref\CombridgeI{B.L. Combridge, J. Kripfganz and J. Ranft, Phys. Lett.
{\bf B70} (1977) 234.}
\nref\Owens{J.F. Owens and E. Reya, Phys. Rev. {\bf D18} (1978) 1501.}
\nref\CombridgeII{B.L. Combridge, Nucl. Phys. {\bf B151} (1979) 429.}
\nref\Morfin{J. Morfin and W.-K. Tung, Z. Phys. {\bf C52} (1991) 13.}
\nref\CTEQ{J. Botts, J. Morfin, J. Owens, J.W. Qui, W.K. Tung and H. Weerts
 MSUHEP-92-27, Fermilab-Pub-92/371, FSU-HEP-92-1225 and ISU-NP-92-17.}
\nref\Harriman{P. Harriman, A. Martin, R. Roberts and J. Stirling, Phys.
 Rev. {\bf D42} (1990) 798.}
\nref\PAKPDF{K. Charchula, Comp. Phys. Comm. {\bf 69} (1992) 360.}
\nref\Gehrels{N. Gehrels, Astro. Phys. J. {\bf 303} (1986) 336.}
\nref\Behrends{S. Behrends and J. Huth, Private communication.}

\nfig\Xsectgraph{Inclusive jet cross section plotted against
transverse jet energy $E_\T$.  The data points are the experimental
measurements reported by CDF.  The solid curve represents the predictions
of pure QCD with no composite interactions, while the dot-dashed curve
illustrates the effect of gluon operator $O_2$ with $\Lambda=2 \TeV$ and
$C_2(\Lambda) = -4 \pi$.  The theoretical results are based upon the
leading order MT set SL distribution function evaluated at $Q^2=E_\T^2/2$.}
\nfig\chisqgraph{$\chi^2$ for 22 degrees of freedom plotted
as a function of $\Lambda^{-2}$. }
%

\def\CITTitle#1#2#3{\nopagenumbers\abstractfont
\hsize=\hstitle\rightline{#1}
\vskip 0.6in\centerline{\titlefont #2} \centerline{\titlefont #3}
\abstractfont\vskip .5in\pageno=0}

\CITTitle{{\baselineskip=12pt plus 1pt minus 1pt
  \vbox{\hbox{CALT-68-1872}\hbox{DOE RESEARCH AND}\hbox{DEVELOPMENT REPORT}
  \hbox{HUTP-93/A018}}}}
  {Looking for Gluon Substructure}{at the Tevatron}
\centerline{Peter Cho\footnote{$^\dagger$}{Work supported in part by the U.S.
 Dept. of Energy under DOE Grant no. DE-FG03-92-ER40701 and by a DuBridge
 Fellowship.}}
\centerline{Lauritsen Laboratory}
\centerline{California Institute of Technology}
\centerline{Pasadena, CA  91125}
\medskip\centerline{and}\medskip
\centerline{Elizabeth Simmons\footnote{$^\ddagger$}{Work supported in
 part by the National Science Foundation under grant no. PHY-9218167 and
the Texas National Research Laboratory Commission under grant no.
RGFY93-278B}}
\centerline{Lyman Laboratory of Physics}
\centerline{Harvard University}
\centerline{Cambridge, MA 02138}

\vskip .3in
\centerline{\bf Abstract}
\bigskip

	The impact of nonrenormalizable gluon operators upon inclusive jet
cross sections is studied.  Such operators could arise in an effective
strong interaction Lagrangian from gluon substructure and would induce
observable cross section deviations from pure QCD at high transverse jet
energies.  Comparison of the theoretical predictions with recent CDF data
yields a lower limit on the gluon compositeness scale $\Lambda$.
We find $\Lambda > 2.03 \TeV$ at $95\%$~CL.

\Date{7/93}

	The inclusive jet cross section data from the 1988-89 Fermilab
Tevatron run span seven orders of magnitude and include the highest
transverse jet energy measurements reported to date \CDF. These data provide
a stringent test of quantum chromodynamics and constrain possible new
physics beyond the Standard Model.  In particular, they set improved limits
on hypothetical quark substructure.  At energies small compared to the
compositeness scale $\Lambda$, the dominant effects from quark substructure
can be reproduced by four-quark operators in a low energy effective
Lagrangian \Eichten.  Although their coefficients are unknown in the absence
of a detailed theory of preon dynamics, the general impact of these
nonrenormalizable operators upon parton scattering may be estimated.
The remarkable agreement between the experimental measurements and
the predictions of QCD then places a bound on the quark compositeness
scale.  CDF finds a lower limit on $\Lambda$ of $1.4 \TeV$ at the $95\%$
confidence level~\CDF.

	In this letter, we reinterpret the CDF data to probe for signals of
new physics that could arise in the gluon sector.  Specifically, we
consider  the impact upon the inclusive cross section measurements  of
nonrenormalizable gluonic operators which may appear in the  effective
strong interaction Lagrangian. Such operators could originate from a number
of  different sources.  For example, suppose there exist new heavy colored
bosons or fermions beyond those in the Standard Model.  Such particles would
induce nonlocal interactions among gluons through loop diagrams.  The leading
behavior of these graphs can readily be  extracted and reexpressed via an
operator product expansion in terms of local but nonrenormalizable gluon
operators.   Alternatively, we might speculate that gluons
are bound states of some more fundamental  preon
constituents.  Then as in the case of composite quarks, preon exchange
could generate nonrenormalizable gluon interactions.  In the following, we
will  adopt a model independent approach and not specify the underlying
physics  whose low energy effects are encoded in the effective Lagrangian.
Instead,  we simply seek to place a limit on its characteristic scale
$\Lambda$.

	We first enumerate the lowest dimension gluon operators
whose scattering effects would be easiest to observe. There exist only two
independent operators of mass dimension $d+2$ in $d=4-\e$ spacetime
dimensions which preserve gauge invariance along with  $C$, $P$ and $T$
\SimmonsI:
\eqn\glueops{\eqalign{
O_1 &= {\muepg \over\Lambda^2} f_{abc} G^\r_{a\v} G^\v_{b\l} G^\l_{c\r} \cr
O_2 &= {1 \over 2!\Lambda^2} D^\r G^a_{\r\v} D_\l G^{\l\v}_a \cr}}
where
\eqn\conventions{\eqalign{
D^\r &= \partial^\r - i\muepg G^\r_a T_a \cr
G^{\r\v}_a &= \partial^\r G^\v_a - \partial^\v G^\r_a + \muepg f_{abc}
G^\r_b G^\v_c \cr}}
and $\u$ denotes the renormalization scale.  All other dimension-$(d+2)$
gluon operators either vanish or reduce to combinations of the two in
eqn.~\glueops.
\foot{The operator obtained from $O_1$ by replacing the antisymmetric
structure constant $f_{abc}$ with the completely symmetric symbol $d_{abc}$
violates charge conjugation and is identically zero.}

	The contributions of $O_1$ and $O_2$ to parton scattering cross
sections have been studied in refs.~\refs{\SimmonsI,\SimmonsII}.
Operator $O_1$ mediates gluon-quark as well as gluon-gluon  scattering and
interferes with the pure QCD amplitudes for these same processes.  One
might expect its impact to become pronounced at higher energies due to the
sizable gluon content of colliding hadrons at small parton momentum fractions.
However, the helicity structure of the $O(1/\Lambda^2)$ amplitude from
$O_1$ for $gg \to gg$ scattering is orthogonal to that of pure QCD.
So the two amplitudes do not interfere.  Similarly, the
$O(1/\Lambda^2)$ interference terms in the differential cross
sections for $gg \to q \qbar$ and the other partonic processes related by
crossing vanish in the limit of zero quark mass.
Thus the presence of $O_1$ in the low energy Lagrangian affects
inclusive jet cross sections starting
only at $O(1/\Lambda^4)$.  This surprising and rather disappointing result
motivates us to search for signals of gluon compositeness in non-gluonic
channels.

	Fortunately, operator $O_2$ mediates quark-quark scattering.
The classical equation of motion
\eqn\EOM{D_\r G^{\r\v}_a = -\muepg \sum_{\rm flavors} \qbar \g^\v T_a q}
relates the S-matrix elements of $O_2$ to those of a color octet
four-quark operator:
\eqn\Smatrixreln{O_2
\> {\buildrel {\scriptscriptstyle EOM} \over \longrightarrow} \>
{\muep g^2\over 2!\Lambda^2} \sum_{\rm flavors}
\bigl( \qbar \g_\v T_a q  \bigr) \bigl( \qbar \g^\v T_a q \bigr).}
This identification should be understood as a relation among S-matrix
elements and not as a true operator identity \Politzer.  Notice
that the equation of motion \EOM\ produces a factor of $g^2$
multiplying the operator in \Smatrixreln.  No such extra powers of the
strong interaction coupling accompany the analogous color singlet operators
that enter into quark compositeness analyses.  At typical
Tevatron energies, the numerical value for $g^2$ is somewhat larger than
unity.  So we anticipate that the effect upon quark scattering from the
four-quark operator induced by gluon substructure will be slightly
larger than that from the four-quark operators generated by quark
substructure.

	To complete our operator basis so that it closes under
renormalization, we include three more four-quark operators
along with $O_1$ and $O_2$ in the effective Lagrangian:
\eqn\effLagrange{\Leff=\CL_\QCD+\sum_{i=1}^5 C_i(\u) O_i(\u)}
where
\eqn\oplist{\eqalign{
O_1 &= {\muepg \over\Lambda^2} f_{abc} G^\r_{a\v} G^\v_{b\l} G^\l_{c\r} \cr
O_2 &= {\muep g^2\over 2!\Lambda^2} \sum_{\rm flavors} \bigl( \qbar \g_\v
 T_a q \bigr) \bigl( \qbar \g^\v T_a q \bigr) \cr
O_3 &= {\muep g^2\over 2!\Lambda^2} \sum_{\rm flavors}
 \bigl( \qbar \g_\v \gfive
T_a  q \bigr) \bigl( \qbar \g^\v \gfive T_a q \bigr) \cr
O_4 &= {\muep g^2\over 2!\Lambda^2} \sum_{\rm flavors} \bigl( \qbar \g_\v q
 \bigr) \bigl( \qbar \g^\v q \bigr) \cr
O_5 &= {\muep g^2\over 2!\Lambda^2} \sum_{\rm flavors} \bigl( \qbar \g_\v
 \gfive q \bigr) \bigl( \qbar \g^\v \gfive q \bigr) . \cr}}
Since the underlying short-distance physics responsible for generating
these nonrenormalizable operators is not known, there
is a fair amount of arbitrariness in the values one chooses for their
dimensionless coefficients and the compositeness scale $\Lambda$.
We will follow the
convention adopted in previous quark substructure studies and define
$\Lambda$ to be the scale where the magnitude of the gluon operator
coefficients $C_1$ or $C_2$ equals $4\pi$ \refs{\Eichten,\EHLQ}.
We set the coefficients of the remaining operators at the $\Lambda$ scale
to zero.

	To determine the operators' coefficients at energies probed
by the Tevatron, we evolve their values down from the compositeness scale
via the renormalization group equation
\eqn\RGE{\mu {d \over d\mu} C_i (\mu) = \sum_j (\gamma^\T)_{ij}
C_j (\mu).}
The anomalous dimension matrix
$$ \g = \bordermatrix{& O_1 & O_2 & O_3 & O_4 & O_5 \cr
O_1 & 7+{2\nf/3} & 0 & 0 & 0 & 0 \cr
O_2 & 0 & 311/36-2\nf/3 & 5/4 & 0 & 2/3 \cr
O_3 & 0 & 41/36 & 35/4-2\nf/3 & 2/3 & 0 \cr
O_4 & 0 & 4/3 & 6 & 11-{2\nf/3} & 0 \cr
O_5 & 0 & 22/3 & 0 & 0 & 11-{2\nf/3} \cr} {g^2 \over 8\pi^2} $$
describes the mixing for $\nf$ active quark flavors among all the
dimension-$(d+2)$ operators in the effective Lagrangian.   The last four
rows of $\g$ quantify the running of the four-quark operators
in our basis and were determined from a straightforward operator mixing
computation.  The entries in the first row on the other hand were
extracted from the highly nontrivial anomalous dimension calculations
reported in ref.~\Narison\ by Narison and Tarrach and in ref.~\Morozov\
by Morozov.  Notice that the triple gluon field strength operator runs only
into itself and does not mix with any of the other four-quark operators.
\foot{As shown in Morozov's paper, the anomalous dimension of
the CP-even $G^3$ operator is the same as that of the CP-odd
$\tilde{G}^3$ operator which gained much notoriety a few years ago
\Braaten.}
Since $O_1$ has virtually no impact upon parton scattering and
does not mix into operators whose effects can be detected, it is
essentially invisible at $O(1/\Lambda^2)$.  We will consequently be unable
to place any limit upon its associated scale.

	It is convenient to decompose the anomalous dimension
matrix as
\eqn\gammadecomp{
\eqalign{\gamma &= {g^2 \over 8\pi^2} \hat \gamma+ O(g^4)  \cr
		   &= {g^2 \over 8\pi^2} S D S^{-1} + O(g^4) \cr}}
where the eigenvalues $\lambda_i$ of $\hat\gamma$ are contained in
the diagonal matrix $D$ while the corresponding eigenvectors are
arranged into the columns of matrix $S$.  The general solution
to the coefficients' renormalization group equation %
\eqn\RGEsolnI{C_i(\mu) = \sum_j \Bigl[ \exp \int^{g(\mu)}_{g(\Lambda)} dg
 {\gamma^\T (g) \over \beta(g)} \Bigr]_{ij} C_j(\Lambda)}
can then be rewritten as
\eqn\RGEsolnII{C_i(\mu) = \sum_{j,k} \bigl(S^{-1} \bigr)^\T_{ij}
 {\rm diag} \Bigl( \Bigl[ {\aS(\mu) \over \aS(\Lambda)}
 \Bigr]^{\lambda_j\over 2b} \Bigr)
 (S^\T)_{jk} \> C_k(\Lambda)}
where $b = -11/2 + \nf /3$ is the coefficient in the one-loop
QCD beta function $\b(g)=b g^3 / 8 \pi^2$ \Gross.
Integrating $\beta$, we obtain the strong interaction fine structure
constant
\eqn\astrong{\aS(\u) = {\aS(\Mz) \over 1-b \displaystyle{\aS(\Mz)\over\pi}
\log{\displaystyle{\u\over\Mz}}}.}
We choose the constant of integration in the integrated beta
function to be $\aS(\Mz)=0.118 \pm 0.007$ \Bethke\ rather than the QCD
scale since parton energies at the Tevatron are typically large compared to
the $Z$ scale.

	We now turn to computing the inclusive jet cross section in $p
\bar{p}$ collisions.  Neglecting higher order multi-jet events, we start
with  the two-jet differential cross section
\eqn\twojetXsect{
{d^3\sigma \over d\eta_1 d\eta_2 dp_\T} \bigl(AB \to 2 \space {\rm jets}
\bigr) = \sum_{abcd} 2 x_a x_b p_\T \Bigl[ f_{a/A}(x_a) f_{b/B}(x_b)
+ (A \leftrightarrow B {\> \rm if \> } a \ne b) \Bigr]
{d\sigma \over d\that}(ab \to cd)}
expressed in terms of the jets' pseudorapidities ($\eta_1, \eta_2$),
their common transverse momentum $(p_\T$), and the momentum fractions
($x_a, x_b$) of partons $a$ and $b$ inside hadrons $A$ and $B$.  The
partons' distribution functions $f_{a/A}(x_a)$ and $f_{b/B}(x_b)$ are folded
together with the differential cross section $d\sigma/d\that$ for the
elementary scattering process $ab \to cd$.
\foot{Note that the Mandelstam invariant $\hat t$ refers to the colliding
partons rather than the incident hadrons.}
The product is then summed over all possible initial and final parton
configurations.  To convert the two-jet expression \twojetXsect\ into an
inclusive single-jet cross section, we integrate over the pseudorapidity
range of one jet, average the other over the pseudorapidity interval
$0.1 \le |\eta| \le 0.7$ visible to the CDF detector, and multiply by two to
count the contributions of both jets to the inclusive cross section:
\eqn\avgXsect{
{1 \over \Delta\eta} \int d\eta {d^2 \sigma \over d\eta dp_\T}
= {2 \over \Delta\eta} \int d\eta_1 \int_{-\infty}^{\infty} d\eta_2
{d^3 \sigma \over d\eta_1 d\eta_2 dp_\T}.}
The result may then be compared with the measurements reported by CDF~\CDF.

	The lowest order QCD predictions for the parton cross sections
\eqn\partonXsect{{d \sigma\over d\that}(ab \to cd) = {\pi \aS^2 \over
\shat^2} \Sigma(ab \to cd)}
have been frequently documented in the literature \refs{\CombridgeI,\Owens}.
They conventionally include initial state color averaging factors and are
written in terms of the partonic invariants $\shat$, $\that$ and $\uhat$.
The QCD formulae for $\Sigma(ab \to cd)$ are modified by the nonrenormalizable
operators in our effective Lagrangian which induce the $O(1/\Lambda^2)$
interference terms tabulated below:
\vfill\eject
\eqna\Sigmaeqn
$$ \eqalignno{
\Sigma(qq'\to qq') &= {4 \over 9} {\shat^2+\uhat^2 \over \that^2}
+{8 \over 9} {(C_2+C_3)\shat^2+(C_2-C_3)\uhat^2 \over \that \Lambda^2}
+ O\Bigl({1 \over \Lambda^4}\Bigr) & \Sigmaeqn a\cr
\Sigma(q\bar{q}\to q' \bar{q}') &= {4\over 9} {\that^2+\uhat^2 \over \shat^2}
+{8 \over 9} {(C_2+C_3)\uhat^2+(C_2-C_3)\that^2 \over \shat \Lambda^2}
+ O\Bigl({1 \over \Lambda^4}\Bigr) & \Sigmaeqn b\cr
\Sigma(qq \to qq) &= {4 \over 9} \Bigl( {\shat^2+\uhat^2 \over \that^2}
+{\shat^2+\that^2 \over \uhat^2} \Bigr)
-{8\over 27} {\shat^2\over \that \uhat} & \cr
&\qquad +{8C_2 \over 9\Lambda^2} \Bigl({\shat^2+\uhat^2 \over \that}
+{\shat^2+\that^2 \over \uhat} \Bigr)
+{8C_3 \over 9\Lambda^2} \Bigl({\shat^2-\uhat^2 \over \that}
+{\shat^2-\that^2 \over \uhat} \Bigr) & \cr
&\qquad+\Bigl({8 (C_2+C_3)\over 27 \Lambda^2}-{16(C_4+C_5)\over 9\Lambda^2}
\Bigr) {\shat^3\over \that \uhat}
+ O\Bigl({1 \over \Lambda^4}\Bigr) & \Sigmaeqn c\cr
\Sigma(q\bar{q}\to q\bar{q}) &= {4\over 9}\Bigl({\shat^2+\uhat^2\over\that^2}
+{\that^2+\uhat^2 \over \shat^2} \Bigr)
-{8\over 27} {\uhat^2\over \shat\that} & \cr
&\qquad +{8C_2 \over 9\Lambda^2} \Bigl({\shat^2+\uhat^2 \over \that}
+{\that^2+\uhat^2 \over \shat} \Bigr)
-{8C_3 \over 9\Lambda^2} \Bigl({\shat^2-\uhat^2 \over \that}
+{\that^2-\uhat^2 \over \shat} \Bigr) &\cr
&\qquad +\Bigl({8 (C_2+C_3)\over 27 \Lambda^2}-{16(C_4+C_5)\over 9\Lambda^2}
\Bigr) {\uhat^3\over \shat\that}
+ O\Bigl({1 \over \Lambda^4}\Bigr) & \Sigmaeqn d\cr
\Sigma(gg\to q\bar{q}) &= {1\over 6}\Bigl({\that^2+\uhat^2 \over \that
\uhat} \Bigr) -{3\over 8} {\that^2+\uhat^2 \over \shat^2}
+ O\Bigl({1 \over \Lambda^4}\Bigr) & \Sigmaeqn e\cr
\Sigma(q\bar{q}\to gg) &= {32\over 27} \Bigl({\that^2+\uhat^2 \over
\that\uhat} \Bigr) -{8\over 3} {\that^2+\uhat^2\over \shat^2}
+ O\Bigl({1 \over \Lambda^4}\Bigr) & \Sigmaeqn f\cr
\Sigma(gq\to gq) &= -{4\over 9}\Bigl({\shat^2+\uhat^2\over\shat\uhat} \Bigr)
+{\shat^2+\uhat^2 \over \that^2}
+ O\Bigl({1 \over \Lambda^4}\Bigr) & \Sigmaeqn g\cr
\Sigma(gg \to gg) &= {9\over 2}\Bigl( 3-{\that\uhat\over \shat^2}
-{\shat\uhat\over \that^2}-{\shat\that\over \uhat^2} \Bigr)
+ O\Bigl({1 \over \Lambda^4}\Bigr). & \Sigmaeqn h\cr
} $$
Here $q'$ denotes a quark not identical in flavor to quark $q$.
We have dropped $O(1/\Lambda^4)$ terms in these formulae since we are
not keeping track of any dimension-$(d+4)$ gluon operators whose
contributions to $\Sigma(ab\to cd)$ are of the same order.  We have also
neglected all parton masses.  At transverse jet energies of a few hundred
GeV, this should be a good approximation except for the top quark whose mass
we assume is $m_t = 140 \GeV$.  Initial state top mass effects may
be safely ignored as the top content of colliding protons and
antiprotons is negligible.  However for processes involving $t
\bar{t}$ production, we replace eqns.~\Sigmaeqn{b}\ and \Sigmaeqn{e}\
with the heavy flavor QCD cross sections given in
ref.~\CombridgeII\ and incorporate $O(1/\Lambda^2)$ interference corrections:
\vfill\eject
\eqna\topeqns
$$ \eqalignno{
\Sigma(q \bar{q} \to t \bar{t}) &= {4 \over 9} {(\that-\mtsq)^2
+(\uhat-\mtsq)^2+2\mtsq \shat \over \shat^2} + {8 \over 9}
{(C_2+C_3)\uhat^2+(C_2-C_3)\that^2+2 C_2 \mtsq\shat \over \shat\Lambda^2}
& \cr
&\qquad + O\Bigl({1 \over \Lambda^4}\Bigr) & \topeqns a \cr
\Sigma(gg \to t \bar{t}) &= {3 \over 4} {(\mtsq-\that)(\mtsq-\uhat) \over
\shat^2} -{1 \over 24} {\mtsq(\shat-4\mtsq) \over
(\mtsq-\that)(\mtsq-\uhat)} & \cr
&\qquad +\sixth{(\mtsq-\that)(\mtsq-\uhat)-2\mtsq(\mtsq+\that) \over
(\mtsq-\that)^2} +\sixth{(\mtsq-\that)(\mtsq-\uhat)-2\mtsq(\mtsq+\uhat)
\over (\mtsq-\uhat)^2} & \cr
&\qquad -{3\over 8}{(\mtsq-\that)(\mtsq-\uhat)+\mtsq(\uhat-\that) \over
\shat(\mtsq-\that)}-{3\over 8}{(\mtsq-\that)(\mtsq-\uhat)+\mtsq(\that-\uhat)
\over \shat(\mtsq-\uhat)} & \cr
&\qquad -{9\over 8} {C_1 \mtsq \over \Lambda^2} \Bigl( 3 + 2 \mtsq
{\that^2+\that\uhat+\uhat^2 \over \shat\that\uhat} - 3 m_t^4 {\that+\uhat
\over \shat \that \uhat} \Bigr)
+ O\Bigl({1 \over \Lambda^4}\Bigr). & \topeqns b \cr} $$
Note that coefficient $C_1$ of the triple gluon field strength operator
enters into eqn.~\topeqns{b}. But since it does not appear in any other
cross section formula at $O(1/\Lambda^2)$, operator $O_1$ has almost no
perceptible effect upon the inclusive cross section. We consequently
ignore it from here on.

	Combining eqns.~\twojetXsect\ -- \topeqns{}, we calculate
the single-jet inclusive cross section as a function of transverse jet
energy $E_\T$.  We perform the computation using the leading order parton
distribution functions of Morfin and Tung (MT set SL) \Morfin\ and the CTEQ
collaboration (CTEQ set L) \CTEQ\ as well as the next-to-leading order
functions of Morfin and Tung (MT sets B1 and S), Harriman, Martin, Roberts and
Stirling (HMRS set B) \Harriman\ and the CTEQ collaboration (CTEQ set MS).
All of these parton distribution functions along with several others are
conveniently contained within the PAKPDF package \PAKPDF.

	Representative results obtained from the MT set SL structure function
evaluated at the renormalization scale $Q^2=E_\T^2/2$ are compared
with the experimental data in \Xsectgraph.  The solid curve in the
figure illustrates the predictions of pure QCD with no nonrenormalizable
operator interactions.  Following the example of the CDF analysis \CDF , we
have multiplied the theoretical  predictions by a normalization factor $n$
to align them with  the data.  A fit for this constant performed over the
region $80 \GeV \le E_\T \le 160 \GeV$ where effects from any compositeness
operator terms are negligible yields $n=1.35 \pm 0.01$.  The resulting
agreement between the shapes of the QCD and experimental cross section
values is  striking.  There is however a slight suggestion of discrepancy
at the  highest measured transverse energies where compositeness operator
effects would be expected to first show up.  We therefore plot in the same
figure the differential  cross sections obtained after setting
$\Lambda=2.0 \TeV$ and $C_2(\Lambda)=- 4\pi$ in our effective Lagrangian.
The resulting dot-dashed curve in \Xsectgraph\ qualitatively appears to fit
the data slightly better.

	To be more quantitative, we perform a least squares fit for the
compositeness scale $\Lambda$.  First we multiply the differential cross
section in each CDF bin by the integrated luminosity and bin-width to
convert into number of events:
\eqn\binevents{N = 4200 \, {\rm nb}^{-1} \times \Bigl( \Delta E_\T \int d\eta
{d^2 \sigma \over d\eta dE_\T} \Bigr) {\rm nb}.}
We then examine the transformed error bars to determine the statistics
obeyed by the binned events.  The statistical uncertainties in each bin
with transverse energy $E_\T < 115 \GeV$ are significantly greater than
$\sqrt{N}$.  We therefore exclude these non-Gaussian points from our least
squares analysis.  Bins in the intermediate energy range $115 \GeV < E_\T <
300 \GeV$ contain large numbers of events and follow Gaussian statistics.
At the highest transverse energies, the data bins have fewer than 25
events and are described by Poisson statistics.  We assign Gaussian error
bars to these last points following the discussion in ref.~\Gehrels\ and
then treat them like Gaussian bins containing greater numbers of events.

	In addition to the statistical fluctuations associated with each bin,
we need to consider the systematic errors.  Normalizing the
theoretical number of events by the factor $n$ removes a large,
correlated systematic uncertainty.  We therefore subtract an averaged
percentage uncertainty from all the bins' error bars. The
corrected systematic errors are then small for bins with transverse
energies above $80 \GeV$.

	We adopt the $\chi^2$ function
\eqn\chisq{\chi^2=\sum_{i,j} \Delta_i (V^{-1})_{ij} \Delta_j}
where $\Delta_i = N_i^{\rm th} - N_i^{\rm exp}$ represents the difference
between the theoretically expected and experimentally measured number of
events in the $i{\rm th}$ bin, while $V$ denotes the covariance matrix.
The statistical and systematic uncertainties for
each bin are summed together in quadrature to form the diagonal entries
$\sigma^2_{ii}=\sigma^2_i(\rm stat) + \sigma^2_i(\rm sys)$
in $V$, while the off-diagonal elements which take into account
residual bin-to-bin correlations among the systematic errors are given by
$\sigma^2_{ij} = \sigma_i(\rm sys) \sigma_j(\rm sys)$ \Behrends.
We illustrate in \chisqgraph\ the dependence of $\chi^2$ for 22 degrees of
freedom upon $\Lambda^{-2}$ over the domain $-0.35 \TeV^{-2} \le \Lambda^{-2}
\le 0.45 \TeV^{-2}$ using the MT set SL structure function.
\foot{Negative values for $\Lambda^{-2}$ do not imply imaginary values for
$\Lambda$.  Instead the phase of $\Lambda^{-2}$ is simply absorbed into
the dimensionless $C_2$ coefficient which multiplies the gluon operator
$O_2$ in our effective Lagrangian.}
Various limits may readily be extracted from the clean parabola appearing
in this plot.  For instance, we find
\eqn\Lambdasqlimits{\Lambda^{-2} = 0.113 \pm 0.080 \TeV^{-2}}
by locating the $\chi^2_{\rm min}+1$ points on the parabola.  This
translates into the asymmetrical $1\sigma$ interval
\eqn\Lambdalimits{\Lambda = 2.98^{\displaystyle +2.59}_{\displaystyle -0.69}
\TeV}
for the gluon compositeness scale.  Alternatively, we may quote the more
conservative lower bound
\eqn\lowerbound{\Lambda > 2.03 \TeV \space {\rm at} \space 95 \% \space
{\rm CL}.}
Analogous results from the other leading and next-to-leading order
distribution functions evaluated at the renormalization scales
$Q^2=E_\T^2/2$ and $Q^2=E_\T^2$ are displayed in table 1.  We see from the
$95\%$ lower limit entries in the last column of this table that the bound
in \lowerbound\ represents a conservative estimate for $\Lambda$.  It
compares favorably with the CDF limit for the compositeness scale
associated with quark substructure.

	It should soon be possible to substantially improve our limit
on new gluon sector physics.  The current 1992-93 Tevatron run is expected
to collect a data sample five times larger than the one used in this analysis.
Cross sections at higher transverse jet  energies will be probed, and
sensitivity to any nonrenormalizable  operators in the effective Lagrangian
will be enhanced.  In the next few  years, an integrated luminosity of
$60 \space {\rm pb}^{-1}$ is projected.  We therefore look forward to
updating our findings as new  data comes forth from Batavia.

\bigskip
\centerline{\bf Acknowledgements}
\bigskip

	We thank Steve Behrends, John Huth, Frank Porter and Mark Wise for
helpful discussions.  We are also grateful to the Aspen Center for Physics
where this work was completed for its warm hospitality.

\vfill\eject


%
$$ \vbox{\offinterlineskip
\def\tablerule{\noalign{\hrule}}
\hrule
\halign {\vrule#& \strut#&
\ \hfil#\hfil& \vrule#&
\ \hfil#\hfil& \vrule#&
\ \hfil#\hfil& \vrule#&
\ \hfil#\hfil& \vrule#&
\ \hfil#\hfil& \vrule#&
\ \hfil#\hfil\ & \vrule# \cr
height10pt && \omit && \omit && \omit && \omit && \omit && \omit &\cr
&& \quad Distribution \quad && $ Q^2 $ && normalization && $\chi^2_{\rm min}$
 && $\Lambda^{-2} \, / \TeV^{-2}$ && $\Lambda_{95} \, / \TeV$ & \cr
&& \quad function \quad && \omit && factor && \omit && \omit && \omit & \cr
height10pt && \omit && \omit && \omit && \omit && \omit && \omit &\cr
\tablerule
height10pt && \omit && \omit && \omit && \omit && \omit && \omit &\cr
&& MT && $ E_\T^2/2 $ && $ 1.35 \pm 0.01 $ && $ 11.76 $ && $ 0.113 \pm
 0.080 $ &&  $ 2.03 $ & \cr
height10pt && \omit && \omit && \omit && \omit && \omit && \omit &\cr
&& set SL && $ E_\T^2 $ && $ 1.61 \pm 0.01 $ && $ 12.17 $ &&
 $ 0.081 \pm 0.067 $ && $ 2.29 $ &\cr
height10pt && \omit && \omit && \omit && \omit && \omit && \omit &\cr
\tablerule
height10pt && \omit && \omit && \omit && \omit && \omit && \omit &\cr
&& MT && $ E_\T^2/2 $ && $ 1.17 \pm 0.01 $ && $ 11.10 $ && $ 0.104 \pm
 0.085 $ &&  $ 2.03 $ & \cr
height10pt && \omit && \omit && \omit && \omit && \omit && \omit &\cr
&& set S && $ E_\T^2 $ && $ 1.39 \pm 0.01 $ && $ 11.34 $ &&
 $ 0.081 \pm 0.071 $ && $ 2.25 $ &\cr
height10pt && \omit && \omit && \omit && \omit && \omit && \omit &\cr
\tablerule
height10pt && \omit && \omit && \omit && \omit && \omit && \omit &\cr
&& MT && $ E_\T^2/2 $ && $ 1.15 \pm 0.01 $ && $ 11.15 $ && $ 0.104 \pm
 0.085 $ &&  $ 2.03 $ & \cr
height10pt && \omit && \omit && \omit && \omit && \omit && \omit &\cr
&& set B1 && $ E_\T^2 $ && $ 1.37 \pm 0.010 $ && $ 11.45 $ &&
 $ 0.081 \pm 0.072 $ && $ 2.24 $ &\cr
height10pt && \omit && \omit && \omit && \omit && \omit && \omit &\cr
\tablerule
height10pt && \omit && \omit && \omit && \omit && \omit && \omit &\cr
&& HMRS && $ E_\T^2/2 $ && $ 1.08 \pm 0.01 $ && $ 11.52 $ && $ 0.081
 \pm 0.092  $ &&  $ 2.08 $ & \cr
height10pt && \omit && \omit && \omit && \omit && \omit && \omit &\cr
&& set B && $ E_\T^2 $ && $ 1.28 \pm 0.01 $ && $ 11.48 $ &&
 $ 0.077 \pm 0.077 $ && $ 2.22 $ &\cr
height10pt && \omit && \omit && \omit && \omit && \omit && \omit &\cr
\tablerule
height10pt && \omit && \omit && \omit && \omit && \omit && \omit &\cr
&& CTEQ && $ E_\T^2/2 $ && $ 1.25 \pm 0.01 $ && $ 15.82 $
 && $ -0.131 \pm 0.070 $  && $ 2.02 $ & \cr
height10pt && \omit && \omit && \omit && \omit && \omit && \omit &\cr
&& set L && $E_\T^2$ && $1.48 \pm 0.01 $ && $ 15.97 $
 && $-0.117 \pm 0.067 $ &&  $ 2.10 $ & \cr
height10pt && \omit && \omit && \omit && \omit && \omit && \omit &\cr
\tablerule
height10pt && \omit && \omit && \omit && \omit && \omit && \omit &\cr
&& CTEQ && $ E_\T^2/2 $ && $ 1.21 \pm 0.01 $ && $ 12.18 $
 && $ 0.014 \pm 0.084 $  && $ 2.45 $ & \cr
height10pt && \omit && \omit && \omit && \omit && \omit && \omit &\cr
&& set MS && $E_\T^2$ && $ 1.43 \pm 0.01 $ && $ 12.22 $
 && $ 0.018 \pm 0.071 $ &&  $ 2.65 $ & \cr
height10pt && \omit && \omit && \omit && \omit && \omit && \omit &\cr
\tablerule }} $$

\centerline{Table 1}

\listrefs
\listfigs
\bye